\documentclass[english,11pt]{article}
\pdfoutput=1
\usepackage[utf8]{inputenc}
\usepackage[T1]{fontenc}
\usepackage{microtype}
\usepackage[usenames,dvipsnames]{color}
\usepackage{epsf,amsmath,amssymb,graphicx,dcolumn,multirow,ifthen}
\usepackage{mathtools}
\usepackage{tikz}
\usetikzlibrary{shapes,arrows,positioning}
\usepackage{scalefnt}
\usepackage{filemod}
\usepackage{pstricks}
\usepackage{cite,hyperref}
\DeclarePairedDelimiter\ceil{\lceil}{\rceil}

\newcommand{\EulerGamma}{\gamma_\text{E}}

\newcommand{\tnnote}[1]{}

\newcommand{\tet}{t^2\langle E(t)\rangle}
\newcommand{\lambdaqcd}{\Lambda_\text{\qcd}}
\newcommand{\dderiv}[3]{\frac{\dd^{#3}#1}{\dd#2^{#3}}}

\newcommand{\mz}{m_Z}

\newcommand{\as}{\alpha_s}
\newcommand{\api}{\frac{\as}{\pi}}
\newcommand{\dd}{\text{d}}
\newcommand{\tbl}[1]{Table\,\ref{#1}}
\newcommand{\sct}[1]{Sect.\,\ref{#1}}
\newcommand{\fig}[1]{Fig.\,\ref{#1}}
\newcommand{\figs}[1]{Figs.\,\ref{#1}}
\newcommand{\eqn}[1]{Eq.\,(\ref{#1})}
\newcommand{\eqns}[1]{Eqs.\,(\ref{#1})}
\newcommand{\noeqn}[1]{(\ref{#1})}
\newcommand{\abbrev}{\scalefont{.9}}
\newcommand{\order}[1]{{\cal O}(#1)}
\newcommand{\lo}{{\abbrev LO}}
\newcommand{\nlo}{{\abbrev NLO}}
\newcommand{\nnlo}{{\abbrev NNLO}}
\newcommand{\ep}{\epsilon}
\newcommand{\citere}[1]{Ref.\cite{#1}}
\newcommand{\citeres}[1]{Refs.\cite{#1}}

\newcommand{\qcd}{{\abbrev QCD}}

\title{\vspace*{-6em}
  \begin{flushright}
    {\small 
      June 2016\\[-1em]
      TTK-16-19
    }
  \end{flushright}
\vspace*{2em} {\bf 
The perturbative QCD gradient flow\\ to three loops
}}
\author{ 
Robert V. Harlander$^1$ and Tobias Neumann$^2$\\[2em]
 {\it $^1$Institute for Theoretical Particle Physics and Cosmology}\\
 {\it RWTH Aachen University, 52066 Aachen, Germany}\\[.5em]
 {\it $^2$University at Buffalo, The State University of New York}\\
 {\it Buffalo, NY 14260-1500, USA}\\[.5em]
 {\small {\tt harlander@physik.rwth-aachen.de}}\\[-.3em]
 {\small {\tt tobiasne@buffalo.edu}}\\[2em]
 \date{}
 }
\begin{document}
\maketitle

\vspace*{1cm}
\begin{abstract}
The gradient flow in \qcd{} is treated perturbatively through
next-to-next-to-leading order in the strong coupling constant. The
evaluation of the relevant momentum and flow-time integrals is
described, including various means of validation. For the vacuum
expectation value of the action density, which turns out to be a useful
quantity in lattice calculations, we find a very well-behaved
perturbative series through \nnlo{}. Quark mass effects are taken into
account through \nlo{}. The theoretical uncertainty due to
renormalization-scale variation is significantly reduced with respect to
\lo{} and \nlo{}, as long as the flow time is smaller than about
0.1\,fm.
\end{abstract}

\vfill

\section{Introduction}\label{sec:intro}

\qcd{} is a remarkable theory in many respects. It has been unchallenged
in the description of strong interactions since its original formulation
of more than 40 years ago~\cite{Fritzsch:1973pi}. It crucially impacts
theoretical descriptions of a vast range of observations, reaching from
the hadron mass spectrum to cross sections at particle colliders.  When
quarks are neglected (quenched \qcd), the only fundamental parameter of
\qcd{} is the strong coupling constant $\as$, which simultaneously
defines a mass scale $\Lambda_\text{\qcd}\sim \order{100\,\text{MeV}}$
due to its momentum dependence implied by quantum field theory
(dimensional transmutation).  Also, due to asymptotic
freedom\cite{Politzer:1973fx,Gross:1973id}, \qcd{} does not have an
ultra-violet cut-off; it remains consistent up to arbitrarily high
energies.

A very successful approach for calculations based on the \qcd{}
Lagrangian is perturbation theory.  It corresponds to an expansion in
the strong coupling constant $\as$ and is applicable for processes
where the typical mass scale $Q$ is much larger than
$\Lambda_\text{\qcd}$. The natural input quantity is therefore
$\as$, whose numerical value at a reference scale $Q_0\gg
\Lambda_\text{\qcd}$ is determined by comparing theoretical predictions
with measurements for known processes. Any dependence of physical
observables on $\Lambda_\text{\qcd}$ is only implicit through $\as$:
\begin{equation}
\begin{split}
\as(Q) \simeq \frac{1}{\beta_0\log(Q^2/\lambdaqcd^2)}\,,
\end{split}
\end{equation}
where $\beta_0$ is the leading order coefficient of the
\qcd{} $\beta$ function and will be defined below.

Perturbation theory is obviously inadequate for calculating observables
like the hadron mass spectrum or the pion decay constant which are
strongly dependent on $\Lambda_\text{\qcd}$. Such problems are
accessible in lattice gauge theory, however\cite{Wilson:1974sk}. In this
approach, space-time is discretized by a characteristic lattice spacing
$a$ which serves as a {\abbrev UV} regulator.

Unfortunately, lattice gauge theory and perturbation theory are not only
complementary approaches to \qcd{}, but there is practically no overlap
region where both would yield competitive results. There is a certain
amount of cross-fertilization though, in particular in the context of
renormalization\,\cite{Martinelli:1994ty,Chetyrkin:1999pq,
  Capitani:1998mq,Dolgov:2002zm} or flavor physics (for a review, see
\citere{Colangelo:2010et}, for example).

A particularly promising theoretical quantity which is accessible both
perturbatively {\it and} on the lattice is the so-called Yang-Mills
gradient
flow\,\cite{Narayanan:2006rf,Luscher:2009eq,Luscher:2010iy,Luscher:2011bx}.
For a particular gauge invariant quantity (the so-called \qcd{} action
density, to be defined in more detail below), it was shown by
L\"uscher\,\cite{Luscher:2010iy} that it exhibits some welcome features
on the lattice which allows its efficient evaluation with rather high
precision. He also explicitly calculated this quantity perturbatively
through next-to-leading order (\nlo{}) and showed that the standard
\qcd{} renormalization of the gauge coupling constant is sufficient in
order to obtain a finite result\cite{Luscher:2010iy}. This property was
later proven to all orders in perturbation theory\cite{Luscher:2011bx}.
The perturbative and the lattice result were found to be compatible over
a significant interval of the so-called flow-time parameter $t$, thus
opening up a wide range of possibilities for cross-fertilization in both
fields.

In lattice \qcd{}, the benefits and the appeal of the gradient flow have
already been established, for example in its use for determining the
absolute mass scale of a lattice
calculation\cite{Luscher:2010iy,Borsanyi:2012zs} (``scale setting'', see
\citere{Sommer:2014mea}, for example). From the perturbative point of
view, on the other hand, the concept has received almost no attention
since the original works of
\citeres{Luscher:2010iy,Luscher:2011bx}. However, since one may expect
rather precise results on the lattice in this framework, this may pose a
challenge for perturbative calculations as well, possibly leading to
interesting first-principle results for \qcd{}.

With this motivation in mind, we are going to study the \qcd{} action
density in the framework of the gradient flow up to next-to-\nlo{}
(\nnlo{}) in a perturbative approach. The perturbative expansion will be
obtained via Wick contractions of the original field operators. This
results in three $D$-dimensional momentum and up to four flow-time
integrations over products of massless Feynman propagators times
exponential factors involving loop momenta and flow-time integration
variables. They are solved by sector decomposition and suitable
numerical integration routines. While quark-mass effects will be
neglected in the \nnlo{} calculation, we show that they can be included
through a simple one-dimensional integral at \nlo{}. 

The remainder of this paper is organized as follows. In the next
section, after briefly recalling the flow-field formalism, the
generation of the perturbative series and the evaluation of the
resulting integrals is described. Additionally, we give a list of checks
performed on our calculation, validating the numerical as well as the
conceptional steps. In Section\,\ref{sec:results}, we present the
numerical value for the \nnlo{} coefficient of the action density, which
is the main result of this paper, and provide a brief analysis of the
numerical effects. We conclude and give a short outlook on possible
extensions and applications of this work in
Section\,\ref{sec:conclusions}.

\section{Formalism}\label{sec:formalism}

\subsection{The flow field}

The theoretical framework of our calculation is defined by the
equations \cite{Luscher:2011bx,Luscher:2010iy}
\begin{equation}
\begin{split}
\partial_t B^a_\mu &= D^{ab}_\nu G^b_{\nu\mu} +
(1-\lambda)\,D^{ab}_\mu\partial_\nu B^b_\nu\,,\qquad B^a_\mu(t=0,x) =
g_0A^a_\mu(x)\,,\\ G^a_{\mu\nu} &= \partial_\mu B^a_\nu - \partial_\nu
B^a_\mu + if^{abc}B^b_\mu B^c_\nu\,,\qquad D^{ab}_\mu =
\delta^{ab}\partial_\mu - if^{abc}B^c_\mu\,,
\label{eq:flow}
\end{split}
\end{equation}
where $B^a_\mu(t,x)$ is the flow field with space-time index $\mu$ and
color index $a$, $g_0$ is the bare \qcd{} coupling constant, $f^{abc}$
are the {\abbrev SU(3)} structure constants, and $A^a_\mu(x)$ is the
fundamental gauge field of \qcd{}. The derivative $\partial_\mu$ is
understood w.r.t.\ the $D$-dimensional Euclidean\footnote{We work in
  Euclidean space in this paper, unless indicated otherwise.} space-time
variable $x$, while $t$ denotes the so-called
flow-time. It is easily seen \cite{Luscher:2010iy} that solutions of
\eqn{eq:flow} for different values of $\lambda$ are related by a
$t$-dependent gauge transformation of the flow field $B_\mu$.

It was shown in \citere{Luscher:2011bx} that the flow equation \noeqn{eq:flow}
can be written as the following integral equation in momentum space:
\begin{equation}
\begin{split}
\tilde B_\mu^a(t,p) \equiv \int\dd^Dx\,e^{-ipx}B^a_\mu(t,x) = 
g_0\,\tilde K_{\mu\nu}(t,p)\tilde A^a_\nu(p) 
+ \int_0^t\dd s\,
\tilde{K}_{\mu\nu}(t-s,p)\tilde R^a_\nu(s,p)\,,
\label{eq:inteq}
\end{split}
\end{equation}
where
\begin{equation}
\begin{split}
  \tilde{K}_{\mu\nu}(t,z) = e^{-tp^2}\delta_{\mu\nu} - \frac{p_\mu
    p_\nu}{p^2} e^{-tp^2}\left(1-e^{\lambda tp^2}\right)\,,
\label{eq:kmunu}
\end{split}
\end{equation}
and 
\begin{multline}
\tilde R_\mu^a(t,p) = \sum_{n=2}^3\frac{1}{n!}\int_{q_1}\cdots
\int_{q_n} (2\pi)^D\delta(p+q_1+\cdots+q_n)\\
\times
X^{(n,0)}(q_1,\ldots,q_n)^{ab_1\cdots b_n}_{\mu\nu_1\cdots \nu_n}
\tilde B_{\nu_1}^{b_1}(t,-q_1)
\cdots
\tilde B_{\nu_n}^{b_n}(t,-q_n)\,,
\label{eq:rnu}
\end{multline}
with $\int_{p} \equiv \int\frac{\mathrm{d}^Dp}{(2\pi)^D}$.
The vertices $X^{(n,0)}$ read
\begin{equation}
\begin{split}
X^{(2,0)}(q,r)^{abc}_{\mu\nu\rho} =&\
if^{abc}\big\{(r-q)_\mu\delta_{\nu\rho}
+2q_\rho\delta_{\mu\nu} - 2r_\nu\delta_{\mu\rho} -
\lambda(q_\nu\delta_{\mu\rho} - r_\rho\delta_{\mu\nu})\big\}\,,\\
X^{(3,0)}(q,r,s)_{\mu\nu\rho\sigma}^{abcd} =&\ 
f^{abe}f^{cde}(
\delta_{\mu\sigma}\delta_{\nu\rho} - \delta_{\mu\rho}\delta_{\sigma\nu})
\\&+ f^{ade}f^{bce}(
\delta_{\mu\rho}\delta_{\nu\sigma} - \delta_{\mu\nu}\delta_{\rho\sigma})
+f^{ace}f^{dbe}(
\delta_{\mu\nu}\delta_{\rho\sigma} - \delta_{\mu\sigma}\delta_{\nu\rho})\,.
\label{eq:vertices}
\end{split}
\end{equation}
The fact that the
first term in \eqn{eq:inteq} is proportional to $g_0$ allows an
iterative solution of that equation, which leads to an asymptotic
series for $B_\mu^a$,
\begin{equation}
\begin{split}
B_\mu^a = \sum_{n\geq 1} g_0^n B_{n,\mu}^a\,.
\label{eq:bexp}
\end{split}
\end{equation}
With each power of $g_0$, the number of fundamental gauge fields
$A_\mu^a$ increases by one. Furthermore, $B_{n,\mu}^a$ involves terms with
$\ceil{n/2},\ceil{n/2}+1,\ldots,n-1$ flow-time integrations, where $\ceil{n/2}$ denotes
the greatest integer less than or equal to $n/2$.

Note that \eqns{eq:kmunu} and \noeqn{eq:vertices} become particularly simple for
$\lambda=0$; for example, the lowest-order solution of the flow-field
equation is simply $\tilde B_\mu^a(t,p) = e^{-tp^2}\tilde A_\mu^a(p)$ in
this case.

\subsection{Calculation of the action density}\label{sec:calc}

The quantity to be computed in this paper is the vacuum expectation
value of the action density,
\begin{equation}
\begin{split}
E(t,x) \equiv \frac{1}{4}\,G^a_{\mu\nu}G^a_{\mu\nu}\,.
\label{eq:actiondensity}
\end{split}
\end{equation}
Since $E(t,x)$ is gauge invariant, we are allowed to set
$\lambda=0$ in our calculation, which minimizes the number of integrals
to be evaluated. The case $\lambda\neq 0$ will be considered as an
important check of our calculation in \sct{sec:checks}.

The perturbative expansion of the vacuum expectation value
\begin{equation}
\begin{split}
\langle E\rangle = 
\frac{1}{2}\langle
 \partial_\mu B_\nu^a\partial_\mu B_\nu^a - 
 \partial_\nu B_\mu^a\partial_\nu B_\mu^a\rangle
+ f^{abc}
\langle (\partial_\mu B_\nu^a) B_\mu^b B_\nu^c\rangle
+ \frac{1}{4} f^{abc}f^{cde}
\langle B_\mu^a B_\nu^b B_\mu^cB_\nu^d\rangle
\label{eq:e}
\end{split}
\end{equation}
is obtained by inserting the asymptotic expansion of the flow field
$B_\mu^a$ as obtained in the previous section, and including higher
orders of the fundamental perturbative vacuum, i.e.
\begin{equation}
\begin{split}
\langle {\cal O}\rangle = \frac{\langle 0| {\cal
    O}\exp(-S_\text{\qcd}(g_0))|0\rangle}{\langle
  0|\exp(-S_\text{\qcd}(g_0))|0\rangle}\,,
\label{eq:pert}
\end{split}
\end{equation}
where $S_\text{\qcd}$ is the interaction part of the fundamental \qcd{}
action which depends on the fundamental gauge fields $A_\mu^a$.  

From \eqns{eq:e} and \noeqn{eq:inteq} it follows that $\langle E\rangle
= \order{g_0^2}$; only the first term on the r.h.s.\ of \eqn{eq:e}
contributes at lowest order, while the second and the third term are of
order $g_0^4$ (odd powers in $g_0$ vanish due to an odd number of fields
in the matrix elements).  However, all three terms on the r.h.s.\ of
\eqn{eq:e} contribute to higher orders as well, either through higher
orders in the expansion of the $B$-fields, see \eqn{eq:inteq}, or
through the perturbative expansion of the exponential in
\eqn{eq:pert}. The former case generally leads to an increase in the
number of flow-time integrations, while the latter corresponds to
corrections due to fundamental \qcd{}. The general form of a matrix
element to be evaluated at order $g_0^{n}$ can therefore be symbolized
by
\begin{equation}
\begin{split}
M_n(k,m) \equiv \langle 0| (B_{m_1}\cdots B_{m_k})\times
\left(S_\text{\qcd}\right)^{n-m}|0\rangle\,,\qquad m=\sum_{i=1}^k m_i\,,
\label{eq:mkm}
\end{split}
\end{equation}
where $B_{m_i}$ is the $m_i^\text{th}$ coefficient of the asymptotic
series in \eqn{eq:bexp}.  This classification turns out useful with
respect to the way we subsequently simplify the matrix elements in the
sense that individual terms cannot be combined among different classes.
Note that $0\leq m-k$ is the maximum number of flow-time integrations in
$M_n(k,m)$, and since $m\leq n$, the maximum number of flow-time
integrations at order $g_0^n$ is\footnote{For a $k$-point function, the
  maximum number of flow-time integrations is $n-k$.}  $n-2$.

One particularly simple class when calculating $\langle E\rangle$ is
$M_n(2,2)$, which is fully determined by the $(n-2)$-loop self-energy of
the fundamental gluon field; this will serve as a welcome check of our
calculation, see \sct{sec:checks}. At \lo{}, $M_2(2,2)$ is in fact the
only class that contributes. At order $g_0^n$ for $n\geq 4$, one needs
to evaluate $3(n-2)$ classes, namely $M_n(k,m)$ with $k\in\{2,3,4\}$ and
$k\leq m\leq n$.  Thus, at \nnlo{}, there are twelve classes that
contribute to $\langle E\rangle$, and the maximum number of flow-time
integrations is four. For comparison, at \nlo{} there are six classes
and at most two flow-time integrations.

\subsection{Evaluation of the perturbative series}

Except for the final numerical integration, all stages of the
calculation were performed with the help of {\tt
  Mathematica}\cite{mathematica70}. Rather than following the
diagrammatic method developed in \citere{Luscher:2011bx}, we directly
implemented the Wick contractions of the gauge and quark fields after
the iterative expansion of the flow fields according to \eqns{eq:inteq}
and (\ref{eq:rnu}), the perturbative expansion of $\exp(-S_\text{\qcd})$
in \eqn{eq:pert}, and the insertion of \qcd{}-Feynman rules. The Dirac
algebra is performed with the functionalities of {\tt
  FeynCalc}\cite{Mertig:1990an} and color factors are calculated using
{\tt ColorMath}\cite{Sjodahl:2012nk}; vacuum diagrams are discarded as
required by the normalization factor in \eqn{eq:pert}.

After these algebraic and symbolic manipulations, one ends up with
integrals of the general form (at $\order{g_0^6}$)
\begin{equation}
\begin{split}
&I(t,{\bf n},{\bf a},D) = \left(\prod_{f=1}^N\int_0^{t_{f}^\text{up}}\dd
  t_f\right)\int_{p_1,p_2,p_3} \frac{\exp[\sum_{k,i,j} a_{kij} t_k
      p_ip_j] }{p_1^{2n_1}p_2^{2n_2}p_3^{2n_3}
    p_4^{2n_4}p_5^{2n_5}p_6^{2n_6}}\,,
\label{eq:i}
\end{split}
\end{equation}
where $D=4-2\ep$ is the space-time dimension,
\begin{equation}
\begin{split}
{\bf n} &= \{n_1,n_2,n_3,n_4,n_5,n_6\}\,,\\
{\bf a} &= \{a_{kij}:k=0,\ldots, N\,;\,i=1,2,3\,;j=1,2,3\}
\end{split}
\end{equation}
are sets of integers, $N\leq 4$, $t_0\equiv t$, and the upper limits for
the flow-time integrations are linear combinations of the other
flow-time variables, $t_f^\text{up}=t_f^{\text{up}}(t_0,\ldots
t_{f-1})$. The momenta $p_4$, $p_5$, $p_6$ are linear combinations of
the integration momenta $p_1$, $p_2$, $p_3$. Quark-mass effects have
been neglected in \eqn{eq:i}; at \nlo{}, we will take them into account
in \sct{sec:quarkmass}.

Needless to say that in order to minimize computer time, it is important
to identify integrals which differ by linear transformations of the loop
momenta and flow-time integration variables at this stage, to cancel
numerators with denominators in the integrals as far as possible, and to
discard scale-less integrals which vanish in dimensional regularization.
After these simplifications, the number of integrals of the form given
in \eqn{eq:i} is listed in \tbl{tab:numbersnnlo}, both split according
to the classification defined in \eqn{eq:mkm}, and according to the
number of flow-time integrations. For comparison, we also give the
corresponding numbers for the \nlo{} case in \tbl{tab:numbersnlo}.

\begin{table}
\begin{center}
\begin{tabular}{c}
\begin{tabular}{c|ccccc|cccc|ccc|c}
$k$&\multicolumn{5}{c|}{2} & 
\multicolumn{4}{c|}{3} & 
\multicolumn{3}{c|}{4} &  \multirow{2}{*}{$\Sigma$}\\
\cline{1-13}
$m$&2 & 3 & 4 & 5 & 6 & 3 & 4 & 5 & 6 & 4 & 5 & 6&\\
\cline{1-14}
\#& 24 & 45 & 219 & 683 & 2244 & 13 & 43 & 110 & 244 & 5 & 7 & 14 & 3651
\end{tabular}\\
(a)\\[2em]
\begin{tabular}{c|ccccc|c}
$f$ &0 & 1 & 2 & 3 & 4 & $\Sigma$\\
\hline
\#& 42 & 117 & 412 & 1229 & 1851 & 3651
\end{tabular}\\
(b)
\end{tabular}
\caption[]{\label{tab:numbersnnlo} Number of integrals at \nnlo{} (a) in
  class $M_6(k,m)$, and (b) involving $f$ flow-time integrations. The
  numbers may not strictly be minimal; they are to be understood as a
  reference, in particular in comparison to the \nlo{} numbers given in
  \tbl{tab:numbersnlo}. }
\end{center}
\end{table}

\begin{table}
\begin{center}
\begin{tabular}[t]{cc}
\begin{tabular}{c|ccc|cc|c|c}
$k$&\multicolumn{3}{c|}{2} & 
\multicolumn{2}{c|}{3} & 
\multicolumn{1}{c|}{4} & \multirow{2}{*}{$\Sigma$}\\
\cline{1-7}
$m$&2 & 3 & 4 & 3 & 4 & 4 & \\
\hline
\#& 1 & 4 & 11 & 1 & 2 & 1 & 20
\end{tabular} &\hspace*{2em}
\begin{tabular}[b]{c|ccc|c}
$f$ &0 & 1 & 2 & $\Sigma$\\
\hline
\#& 3 & 7 & 10 & 20
\end{tabular}\\
(a) &\hspace*{2em} (b)
\end{tabular}
\caption[]{\label{tab:numbersnlo} Number of integrals at \nlo{} (a) in
  class $M_4(k,m)$, and (b) involving $f$ flow-time integrations. The
  numbers may not strictly be minimal; they are to be understood as a
  reference, in particular in comparison to the \nnlo{} numbers given in
  \tbl{tab:numbersnnlo}.}
\end{center}
\end{table}

When quark masses are neglected, the only mass scale in the problem is
the flow time $t$, and therefore
\begin{equation}
\begin{split}
I(t,{\bf n},{\bf a},D) = t^{-d/2}\,c({\bf n},{\bf a},D)\,,\qquad
d=3D-2N-2\sum_{i=1}^6 n_i\,,
\label{eq:ic}
\end{split}
\end{equation}
where $c({\bf n},{\bf a},D)$ is dimensionless.

Introducing Schwinger parameters as
\begin{equation}
\begin{split}
\frac{1}{p^{2n}} &=
\frac{1}{(n-1)!}\int_0^\infty \dd s\,s^{n-1}\, e^{-sp^2}\,,\qquad
p^{2n} = \dderiv{}{s}{n} e^{sp^2}\bigg|_{s=1}\,,
\end{split}
\end{equation}
where $n\in\mathbb{N}$, the momentum integration reduces to a Gaussian
integral:
\begin{equation}
\begin{split}
\int_{p_1,p_2,p_3} \exp[-{\bf p}^T A({\bf s},{\bf t}) {\bf p}] = 
\left(\det A({\bf s},{\bf t}) \right)^{-D/2} \left( 4\pi \right)^{-3D/2}\,,
\end{split}
\end{equation}
where ${\bf p}=(p_1,p_2,p_3)$, and $A({\bf s},{\bf t})$ is a coefficient
matrix which is linear in the Schwinger parameters ${\bf s} =
\{s_1,\ldots,s_6\}$ and the flow-time variables ${\bf
  t}=\{t_0,\ldots,t_N\}$.

Through simple rescaling of the flow-time variables and the Schwinger
parameters,
\begin{equation}
\begin{split}
t_n \to \frac{t_n}{t_n^\text{up}}\,,\qquad
s_n \to \frac{s_n}{s_n-1}\,,
\end{split}
\end{equation}
one ends up with integrals of the form
\begin{equation}
\begin{split}
J(D) = \int_0^1\dd x_1\cdots \int_0^1\dd x_M
  \prod_i P_i^{a_i}(x_1,\ldots,x_M)\,,
\label{eq:jtd}
\end{split}
\end{equation}
where $M>0$, the $P_i$ are polynomials in $x_1,\ldots, x_M$, and the
exponents $a_i$ can be $D$-dependent. In the limit $4-D=2\ep\to 0$, the
integrals develop divergences. The integration over the $x_n$ can be
carried out analytically only for a few simple cases, which is why one
needs to resort to numerical integration.\footnote{For attempts of
  analytically evaluating the three-loop integrals, see
  \citere{diss:neumann}.}  This requires the isolation of the terms that
become singular as $\ep\to 0$, which can be achieved algorithmically
through sector decomposition\cite{Binoth:2003ak}. In our calculation, we
apply this method through the {\tt Mathematica} package {\tt
  FIESTA}\cite{Smirnov:2013eza}, which provides us with the result in
the form
\begin{equation}
\begin{split}
J(D) = \frac{1}{\ep^2}J_2 + \frac{1}{\ep}J_1 +
J_0 + \ldots\,,
\end{split}
\end{equation}
where the ellipsis denotes higher order terms in $\ep=(4-D)/2$, and the
$J_n$ are convergent integrals over rational functions times
logarithms of the parameters $x_1,\ldots, x_M$. They can thus be
evaluated numerically.  We prevented {\tt FIESTA} from performing this
integration, and rather used a fully symmetric integration rule of order
13 \cite{genz}.  All parts of the integration are performed with high
precision arithmetics using the {\abbrev MPFR}
library.\footnote{\url{http://www.holoborodko.com/pavel/mpfr/},
  \url{http://www.mpfr.org/}} We checked that this algorithm provides
us with a reliable estimate of the numerical accuracy.

Let us give the explicit result for one particular non-trivial integral
of the type in \eqn{eq:i} which occurs in the calculation of $\tet$.  It
has four flow-time integrations and thus belongs to the class $M_6(2,6)$.
Furthermore, from the flow-time integration limits, we see that it
originates from the iterated insertion of four 3-point flow-time vertices
$X^{(2,0)}$:
\begin{multline}
\int_{k,q,r}
\int_0^t\mathrm{d}s_0
\int_0^{s_0}\mathrm{d}s_1
\int_0^{s_1}\mathrm{d}s_2
\int_0^{s_2}\mathrm{d}s_3\,
\frac{(k+q)^2 (k+r)^2}{(k-q)^2 (q-r)^2}\times\\
\exp\big[2r(r-q)(s_0+s_3)+2kr(s_0-s_1) 
+2kq(s_1-s_2-2t) +2k^2t+2q^2(s_2+t)\big]\\
= \frac{t^{-2+3\epsilon}}{(4\pi)^{3D/2}}
\left(-0.858906438(2)
+\frac{0.0078125}{\epsilon^2}
-\frac{0.0037791975(3)}{\epsilon}\right)\,.
\label{eq:sampleintegral}
\end{multline}
The numerical result in the last line is obtained by following the
evaluation procedure described above. The numbers in brackets indicate
the integration error; for the $1/\ep^2$-terms we were able to derive an
analytical result, for which we simply quote the first few digits of its
numerical value. The precision of order $10^{-9}$ as quoted in
\eqn{eq:sampleintegral} for the $1/\ep^0$-term corresponds to about
250\,{\abbrev CPU} minutes on an 3\,GHz {\abbrev AMD A}8 processor;
a precision of $10^{-6}$ ($10^{-4}$) could be achieved within about ten
(two) minutes. The {\abbrev CPU} time for the $1/\ep$-term is typically
several orders of magnitude smaller.

\subsection{Validation of the calculation}\label{sec:checks}

Since this is the first three-loop calculation in the gradient-flow
formalism, we considered it of utmost importance to validate our setup.
We successfully completed the following checks.

\paragraph{Lower order results.}

It is important to note that our calculation does not rely on any of the
results of \citeres{Luscher:2010iy,Luscher:2011bx}. The fact that we reproduced
the \nlo{} results evaluated in these papers is therefore an important
check of the setup in general. Since the \nlo{} result is known
analytically, we can use it also to cross check the numerical accuracy
claimed by our integration routine, and we find rather conservative
estimates. Specifically, our numerical result agrees with the analytical
expression through $10^{-15}$.

\paragraph{UV-poles at \nnlo.}

The terms of order $1/\ep^2$ and $1/\ep$ obtained in our three-loop
calculation need to be cancelled by the corresponding terms due to the
renormalization of the strong coupling constant at lower orders. We
verify this cancellation by analytical integration for the $1/\ep^2$
terms, and numerically through one part in $10^{10}$ for the $1/\ep$
terms.
Note that the number and complexity of the integrals is typically
smaller for higher order poles. However, even though this means that we
cannot expect the same numerical accuracy for the finite terms, it
should still be sufficient for any foreseeable practical application.

We note in passing that, in the case of the quantity under
consideration, the cancellation of the poles is equivalent to the
renormalization group ({\abbrev RG}) invariance of the final result:
\begin{equation}
\begin{split}
\mu^2\frac{\dd}{\dd \mu^2}\langle E(t)\rangle = 0\,,
\label{eq:rginv}
\end{split}
\end{equation}
where $\mu$ is the renormalization scale. The quantity $\langle
E\rangle$ depends on $\mu$ implicitly through $\as(\mu)$, and
explicitly through terms of the form $\ln \mu^2\,t$. Knowing the
logarithmic dependence in $t$ is thus equivalent to knowing the one in
$\mu$. The former is directly obtained from expanding \eqn{eq:ic} for
$\ep\to 0$, while the latter follows from {\abbrev RG}-invariance and
can be derived from lower order terms through the perturbative solution
of the \qcd{} renormalization group equation:
\begin{equation}
\begin{split}
&\mu^2\dderiv{}{{\mu^2}}{}\as(\mu) =
  \as(\mu)\beta(\as)\,,\qquad \beta(\as) = -\sum_{n\geq
    0}\beta_n\left(\frac{\as}{\pi}\right)^{n+1}\,,
\label{eq:rgeq}
\end{split}
\end{equation}
\begin{equation}
\begin{split}
\Rightarrow\qquad
  & \as(q) = \as(\mu) \left[1 +
    \frac{\as(\mu)}{\pi}\beta_0\ln\frac{\mu^2}{q^2} +
    \left(\frac{\as(\mu)}{\pi}\right)^2\left[
      \beta_1\ln\frac{\mu^2}{q^2} 
      + \beta_0^2\ln^2\frac{\mu^2}{q^2}\right] + \ldots\right]\,,
\label{eq:amu}
\end{split}
\end{equation}
with the first two coefficients of the $\beta$ function given
by\footnote{We quote only the \qcd{} $\beta$ function here. The
  coefficients for a general Lie group can be found in
  \citere{vanRitbergen:1997va,Czakon:2004bu}, for example.}
\begin{equation}
\begin{split}
\beta_0 = \frac{11}{4}-\frac{1}{6}n_f\,,\qquad
\beta_1 = \frac{51}{8}-\frac{19}{24}n_f\,,
\label{eq:beta}
\end{split}
\end{equation}
where $n_f$ is the number of active quark flavors.

\paragraph{Two-loop gluon propagator.}

As already pointed out above (see the discussion after \eqn{eq:mkm}),
the class $M_n(2,2)$, where in the first term on the r.h.s.\ of
\eqn{eq:e} the flow fields $B^a_\mu$ in \eqn{eq:e} are replaced by
their lowest-order terms $B^a_{1,\mu}$, is fully determined by the
fundamental gluon self-energy. In fact, using Feynman
gauge and adopting the notation of \citere{Luscher:2010iy}, we may write
\begin{equation}
\begin{split}
{\cal E}_0 \equiv \frac{g_0^2}{2}\langle \partial_\mu
B_{1,\nu}^a\partial_\mu B_{1,\nu}^a - \partial_\nu
B_{1,\mu}^a\partial_\nu B_{1,\mu}^a\rangle = 4g_0^2(D-1)\int_p
\frac{e^{-2tp^2}}{1-\omega(p)}\,,
\label{eq:e0}
\end{split}
\end{equation}
with the gluon self-energy
\begin{equation}
\begin{split}
\omega(p) = \sum_{k=1}^\infty g_0^{2k}(p^2)^{-k\ep}\,
	\frac{\tilde{\omega}_k\, e^{-k\epsilon\EulerGamma}}{(4\pi)^{kD/2}}\,.
\label{eq:omega}
\end{split}
\end{equation}
Using
\begin{equation}
\begin{split}
\int_p e^{-2tp^2} (p^2)^{-k\ep} =
\frac{(2t)^{k\ep}}{(8\pi t)^{D/2}}\frac{\Gamma(D/2-k\ep)}{\Gamma(D/2)}\,,
\end{split}
\end{equation}
the perturbative expansion of \eqn{eq:e0} can be calculated
analytically.  The coefficients $\tilde{\omega}_i$ can be taken from the
literature.\footnote{Two-loop calculations of the gluon propagator for
  were first reported in
  \citeres{Caswell:1974gg,Jones:1974mm,Tarasov:1976ef,Egorian:1978zx};
  we use the result quoted in \citere{Davydychev:1997vh} here.} In
Feynman gauge, they read
\begin{equation}
\begin{split}
\tilde{\omega}_1 &= C_A \left( \frac{5}{3\epsilon} + \frac{31}{9}
\right) - n_f T_R \left( \frac{4}{3\epsilon} +\frac{20}{9} \right) +
\mathcal{O}(\epsilon) \,,\\ \tilde{\omega}_2 &= - C_A^2 \left(
\frac{25}{12\epsilon^2} + \frac{583}{72\epsilon} + \frac{14311}{432} -
\zeta(3) - \frac{25}{12}\zeta(2) \right)\\ &\phantom{=} + 2n_f C_F T_R
\left( \frac{1}{\epsilon} + \frac{55}{6} -
8\zeta(3)\right)\\ &\phantom{=} + 2 n_f C_A T_R \left(
\frac{5}{6\epsilon^2} + \frac{101}{36\epsilon} + \frac{1961}{216} +
4\zeta(3) - \frac{5}{6}\zeta(2) \right) + \mathcal{O}(\epsilon)\,,
\label{eq:omega12}
\end{split}
\end{equation}
where $C_A$ and $C_F$ are the Casimir operators of the adjoint and the
fundamental representation of the underlying gauge group, $T_R$ is the
corresponding trace normalization (in \qcd{}, $C_A=3$, $C_F=4/3$, and
$T_R=1/2$), and $\zeta(z)$ is Riemann's zeta function with the values
$\zeta(2) = \pi^2/6=1.64493\ldots$ and $\zeta(3) =
1.20206\ldots$. Inserting them via \eqn{eq:omega} into \eqn{eq:e0}, we
can compare the result for ${\cal E}_0$ obtained in this way with our
completely independent evaluation which follows the procedure described
in \sct{sec:calc}.  We find agreement at the level of one part in
$10^{8}$.

\paragraph{Derivatives in the flow time.}

Given an integral of the form $I(t,{\bf a},{\bf n},D)$ in \eqn{eq:i}, we
can compute the derivative w.r.t.\ $t$ in two ways: either by applying
it to the integrand on the l.h.s.\ of \eqn{eq:i} and then
calculating the resulting integrals with our setup, or by using
\eqn{eq:ic}, which implies
\begin{equation}
\begin{split}
t\dderiv{}{t}{}I(t,{\bf a},{\bf n},D) = -\frac{d}{2}\,I(t,{\bf a},{\bf n},D)\,,
\end{split}
\end{equation}
with $d$ given in \eqn{eq:ic}. We have confirmed the equivalence of both
approaches in our setup for some of the most complicated integrals at
the level of one part in $10^{10}$.

\paragraph{Gauge parameter independence.} Our setup allows us in
principle to perform the calculation for arbitrary gauge parameter
$\lambda\neq 0$, see \eqn{eq:flow}. We have confirmed general
$\lambda$-independence at \nlo{}, where the number of terms to be
evaluated increases by about a factor of ten compared to the case
$\lambda=0$. At \nnlo{}, however, the sheer volume of integrals when
allowing for general $\lambda$ makes it impossible to evaluate all of
them with meaningful precision in reasonable time.  A much more
practical though still powerful way is to perform an expansion around
$\lambda=0$ and consider only the terms linear in $\lambda$.  The most
significant simplification following from this is that instead of
\eqn{eq:kmunu}, we obtain
\begin{equation}
\begin{split}
\tilde{K}_{\mu\nu}(t,z) \approx e^{-tp^2}\left(\delta_{\mu\nu} + t\lambda
p_\mu p_\nu\right)\,.
\end{split}
\end{equation}
In this way, the number of integrals increases again only by a factor of
$\order{10}$ relative to the case $\lambda=0$. We find gauge parameter
independence of the \nnlo{} result for $\langle E\rangle$ at
$\order{\lambda}$ through $10^{-3}$ for the finite term, and $10^{-10}$
for the $1/\ep$ pole terms.

\subsection{\nlo{} quark-mass effects}\label{sec:quarkmass}

\begin{figure}[h]
    \centering\includegraphics[]{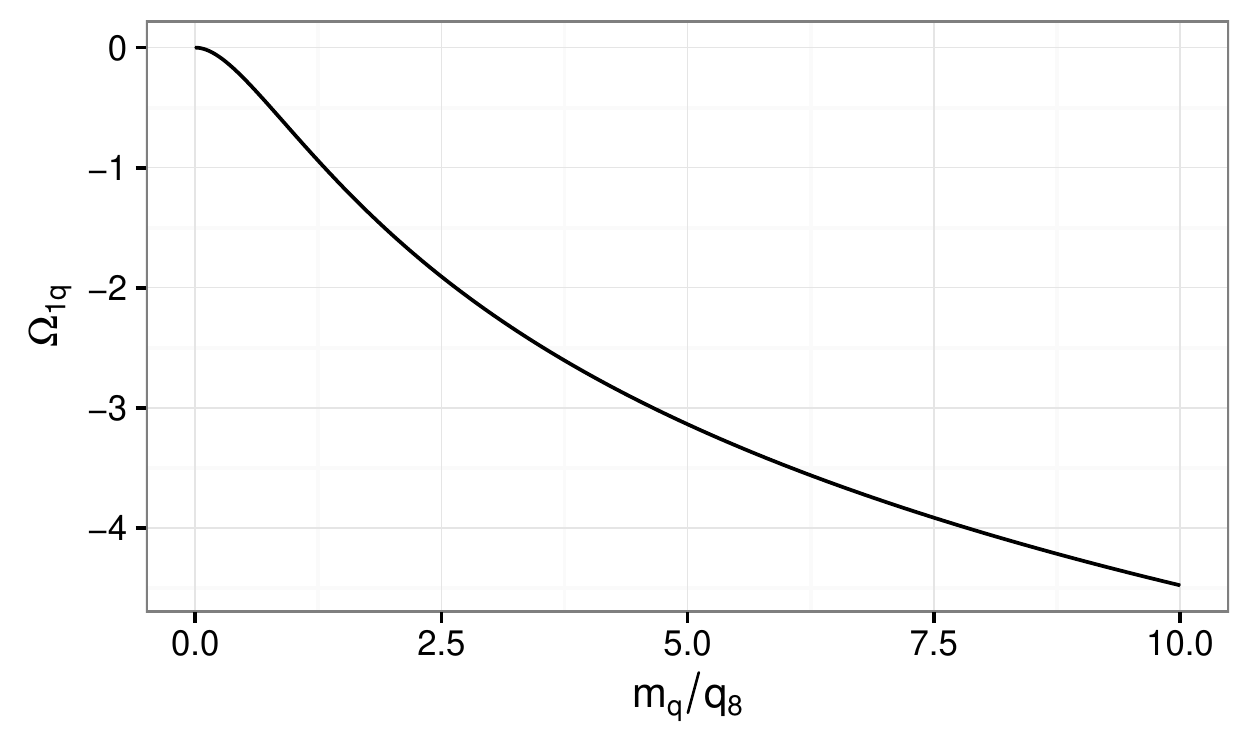}
    \caption{Quark-mass effects $\Omega_{1q}(m_q/q_8)$ as in \eqn{eq:omega1q}.}
    \label{fig:omegaq}
\end{figure}

Quark loops occur first through the one-loop gluon self-energy,
\eqn{eq:omega}. Quark-mass effects can therefore be taken into account
along the lines of \eqns{eq:e0}--\noeqn{eq:omega12} by replacing
\begin{equation}
  \begin{split}
    \tilde \omega_1\to \tilde\omega_1 + \sum_q\Delta\tilde\omega_{1q}\,,
  \end{split}
\end{equation}
where the sum runs over all active quark flavors $q$, and the quark-mass
($m_q$) terms are
given by \cite{Kallen:1955fb}
\begin{equation}
  \begin{split}
    \Delta\tilde\omega_{1q} &=
    \frac{4}{3}\ln\frac{\mu^2}{m_q^2} - \frac{4}{3z_q}
    + \frac{8(1+z_q)(1-2z_q)}{3z_q} \frac{u_q\ln u_q}{u_q^2-1}\,,\\
    z_q &= \frac{p^2}{4m_q^2}\,,\qquad
       u_q = \frac{\sqrt{1+1/z_q}-1}{\sqrt{1+1/z_q}+1}\,.
  \end{split}
\end{equation}
We thus find
\begin{equation}
  \begin{split}
    \tet = \tet\bigg|_{m_q=0} - \frac{\alpha_s^2}{8\pi^2}\sum_q
    \Omega_{1q}\,,
  \end{split}
\end{equation}
where
\begin{equation}
  \begin{split}
\Omega_{1q} = 1 - \EulerGamma - \ln2tm_q^2
- 8m_q^2t+
32t^2m_q^2\int_0^\infty\dd p^2\,e^{-2tp^2}
(1+z_q)(1-2z_q) \frac{u_q\ln u_q}{u_q^2-1}\,.
\label{eq:omega1q}
  \end{split}
\end{equation}
The function $\Omega_{1q}$ depends only on $8m_q^2t\equiv m_q^2/q_8^2$;
its numerical size is displayed in \fig{fig:omegaq}. In the limits of
small and large quark mass, one finds
\begin{equation}
  \begin{split}
    \Omega_{q1}\quad\to\quad
    \left\{
    \begin{array}{l}
      -12\, m_q^2t + \order{(m_q^2t)^2}\,,\\[1em]
      -\ln 2m_q^2t - \EulerGamma - \frac{2}{3} + \order{(m_q^{2}t)^{-1}}\,.
    \end{array}
    \right.
  \end{split}
\end{equation}

\section{Results}\label{sec:results}

\subsection{Action density at three-loop level}\label{sec:nnlo}

We write the result for the vacuum expectation value of the action
density as
\begin{equation}
\begin{split}
\langle E(t)\rangle = \frac{3\as}{4\pi
  t^2}\,\frac{N_A}{8}\,K_E(\as)\,,
\label{eq:nnlotet}
\end{split}
\end{equation}
with the \nnlo{} correction factor
\begin{equation}
\begin{split}
K_E(\as) &= 1 + \as\,k_1
 + \as^2\, k_2\,,
\label{eq:nnloke}
\end{split}
\end{equation}
where $\as\equiv \as^{(n_f)}(\mu)$ is the strong coupling
renormalized at the scale $\mu$ with $n_f$ active quark flavors (assumed
massless), and $N_A$ is the dimension of the adjoint representation of
the underlying gauge group ($N_A=8$ in \qcd{}). Setting
$\mu=1/\sqrt{8t}$, the perturbative coefficients read
\begin{equation}
\begin{split}
  k_1 & =  8\cdot(0.045741114\, C_A +  0.001888798 \, T_R n_f)
  - T_R\sum_q\frac{\Omega_{1q}}{3\pi}\\[-1em]
  &\stackrel{\text\qcd{}}{\approx} 1.098 + 0.008\,n_f + \order{m_q^2t}\,,\\[1em]
	k_2 & = 8\cdot(-0.0136423(7)\, C_A^2 \\
		& \quad + T_R n_f\, \left( 0.006440134(5) \, C_F - 0.0086884(2) \, C_A \right)\\
	& \quad + T_R^2 n_f^2\, 0.000936117)\\
        &\stackrel{\text\qcd}{\approx}
        -0.982 - 0.070\,n_f + 0.002\,n_f^2\,.
\label{eq:k1k2}
\end{split}
\end{equation}
The \nlo{} coefficient $k_1$ has been obtained analytically for $m_q=0$
in \citere{Luscher:2010iy}; we add mass effects $\Omega_{1q}$ obtained
in \eqn{eq:omega1q}. However, for most of our analysis, we find that
these terms are numerically irrelevant, and we will neglect them unless
stated otherwise.  The \nnlo{} coefficient $k_2$ is the main result of
our paper. Similar to \eqn{eq:sampleintegral}, the numbers in brackets
denote the numerical uncertainty. The $n_f^2$-term in $k_2$ is
completely determined by the two-loop gluon propagator, given
analytically in \eqn{eq:omega}. Similar to $k_1$, we simply quote the
first few digits of its numerical value.  Although our main focus is on
\qcd{}, we expressed the result of Eqs.\,(\ref{eq:nnloke},\ref{eq:k1k2})
in terms of ``color'' factors of a general simple Lie group (see above).
For illustration, we also inserted their \qcd{} values and find very
well-behaved perturbative coefficients for any realistic value of $n_f$.

The expression of $\tet$ for general values of the renormalization scale
$\mu$ is easily reconstructed using \eqn{eq:amu}.  \fig{fig:mudep} shows
the variation with this unphysical scale for various values of
\begin{equation}
\begin{split}
q_8\equiv 1/\sqrt{8t}\,.
\end{split}
\end{equation}
From the input value
$\as^{(5)}(\mz)=0.118$, we proceed as described in \fig{fig:running} in
order to derive $\as^{(n_f)}(q_8)$. Here, $l$-loop running of $\as$
means that we numerically solve \eqn{eq:amu} including the coefficients
$\beta_0,\ldots,\beta_{l-1}$. The decoupling of heavy quarks is
consistently performed at $(l-1)$-loop order at the matching scales
$\mu_b=m_b=4.78$\,GeV for $\as^{(5)}\to \as^{(4)}$, and
$\mu_b=2m_c=2\cdot 1.67$\,GeV for $\as^{(4)}\to \as^{(3)}$ (see
\citeres{Chetyrkin:2000yt,Harlander:2002ur} for more details, for
example). These start values are then further evolved for fixed $n_f$ at
the corresponding loop order in order to produce the plots: for the
\lo{}/\nlo{}/\nnlo{} result, we apply one/two/three-loop running of
$\as$.

\tikzstyle{block} = [rectangle, draw, fill=white, text centered, rounded
  corners, minimum height=2em] \tikzstyle{line} = [draw, -latex']
\begin{figure}
\begin{center}
\begin{tikzpicture}[align=center,node distance=2cm]
    \node [block,thick] (mz-5) {$\alpha^{(5)}_s(\mz)=0.1180$};
    \node [block] (100-5) [above=2em of mz-5] {$\alpha^{(5)}_s(100\,\text{GeV})=0.1164$};
    \path [line] (mz-5)-- node [right] {$$} (100-5);
    \node [block] (10-5) [right=4em of mz-5]
          {$\alpha^{(5)}_s(5\,\text{GeV})=0.2131$};
    \path [line] (mz-5)-- node [below] {$$} (10-5);
    \node [block] (mb-5) [below=2em of mz-5]
          {$\alpha^{(5)}_s(m_b)=0.2159$};
    \path [line] (mz-5)-- node [right] {$$} (mb-5);
    \node [block] (mb-4) [below=2em of mb-5]
          {$\alpha^{(4)}_s(m_b)=0.2153$};
    \path [line,dashed] (mb-5)-- node [right] {$$} (mb-4);
    \node [block] (3-4) [right=4em of mb-4]
          {$\alpha^{(4)}_s(4\,\text{GeV})=0.2284$};
    \path [line] (mb-4)-- node [below] {$$} (3-4);
    \node [block] (mc-4) [below=2em of mb-4]
          {$\alpha^{(4)}_s(2m_c)=0.2436$};
    \path [line] (mb-4)-- node [right] {$$} (mc-4);
    \node [block] (mc-3) [below=2em of mc-4]
          {$\alpha^{(3)}_s(2m_c)=0.2360$};
    \path [line,dashed] (mc-4)-- node [right] {$$} (mc-3);
    \node [block] (1-3) [right=4em of mc-3]
          {$\alpha^{(3)}_s(\frac{1}{1.6}\,\text{GeV})=1.415$};
    \path [line] (mc-3)-- node [below] {$$} (1-3);
    \node [block] (15-3) [left=4em of mc-3]
          {$\alpha^{(3)}_s(1\,\text{GeV})=0.4866$};
    \path [line] (mc-3)-- node [below] {$$} (15-3);
    \node [block] (100-3) [below left=2em and -2em of mc-3]
          {$\alpha^{(3)}_s(\frac{1}{1.3}\,\text{GeV})=0.6886$};
    \path [line] (mc-3)-- node [right] {$$} (100-3);
    \node [block] (100-3) [below right=2em and -2em of mc-3]
          {$\alpha^{(3)}_s(\frac{1}{1.5}\,\text{GeV})=0.9848$};
    \node [block] (a1-3) [above=1.5em of 15-3]
          {$\alpha^{(3)}_s(10\,\text{GeV})=0.1662$};
    \node [block] (a2-3) [above=1.5em of a1-3]
          {$\alpha^{(3)}_s(100\,\text{GeV})=0.1043$};
    \path [line] (mc-3)-- node [right] {$$} (100-3);
    \path [line] (mc-3)-- node [right] {$$} (a1-3);
    \path [line] (mc-3)-- node [right] {$$} (a2-3);
\end{tikzpicture}
\end{center}
\caption[]{\label{fig:running} Evolution of $\as$ from the input
  value $\as^{(5)}(\mz)$. Solid arrows denote four-loop {\abbrev
    RG} evolution, dashed arrows three-loop decoupling of heavy quarks.}
\end{figure}
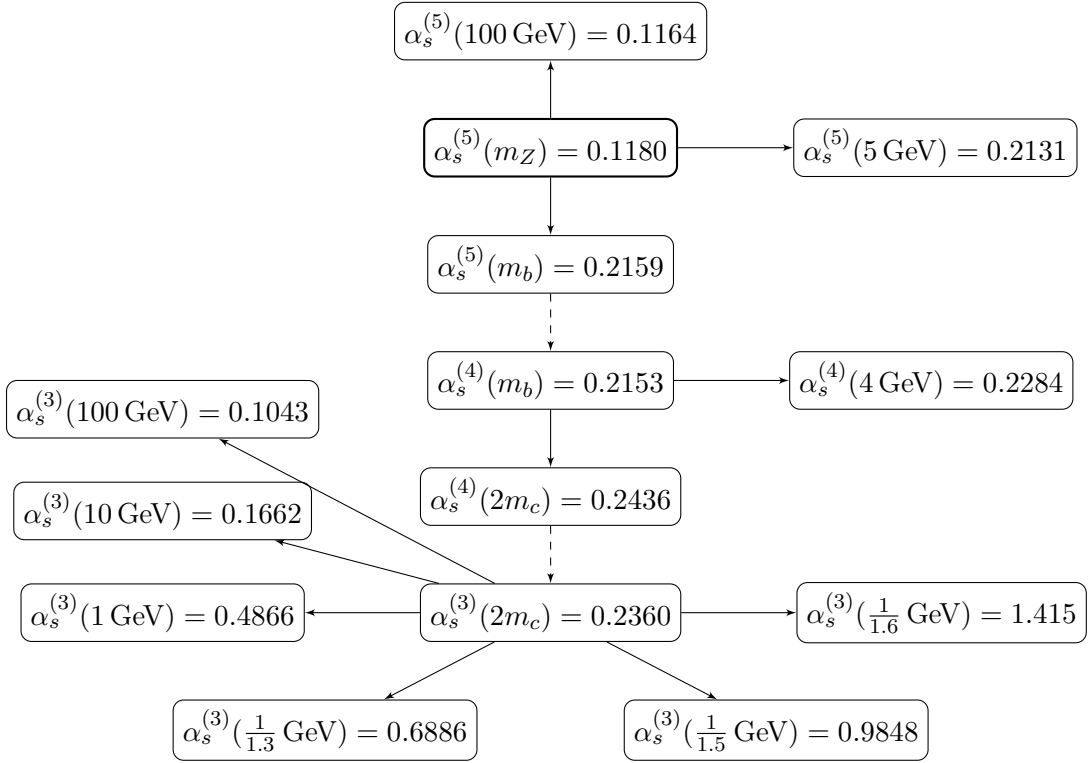

\begin{figure}
  \begin{center}
      \includegraphics{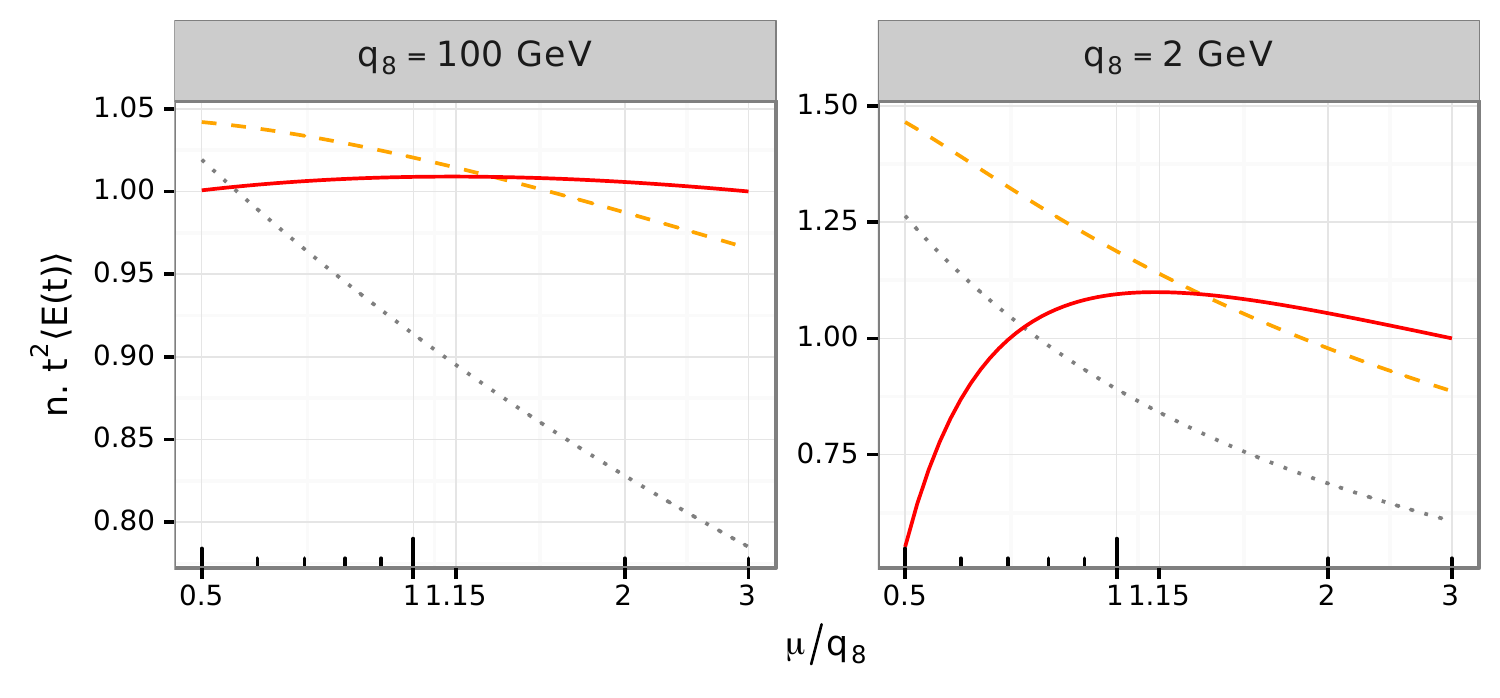}
    \parbox{.9\textwidth}{
      \caption[]{\label{fig:mudep2}\sloppy $\tet$ for $n_f=3$ as a function of
        $\mu/q_8$ for $q_8=100$\,GeV and $q_8=2$\,GeV at \lo{} (black
        dotted), \nlo{} (orange dashed), and \nnlo{} (red solid). All
        curves are normalized to the \nnlo{}-result at $\mu=3q_8$. Note
        the different scales in the two plots. }}
  \end{center}
\end{figure}

\begin{figure}
\hspace{-1.5em}
	\includegraphics{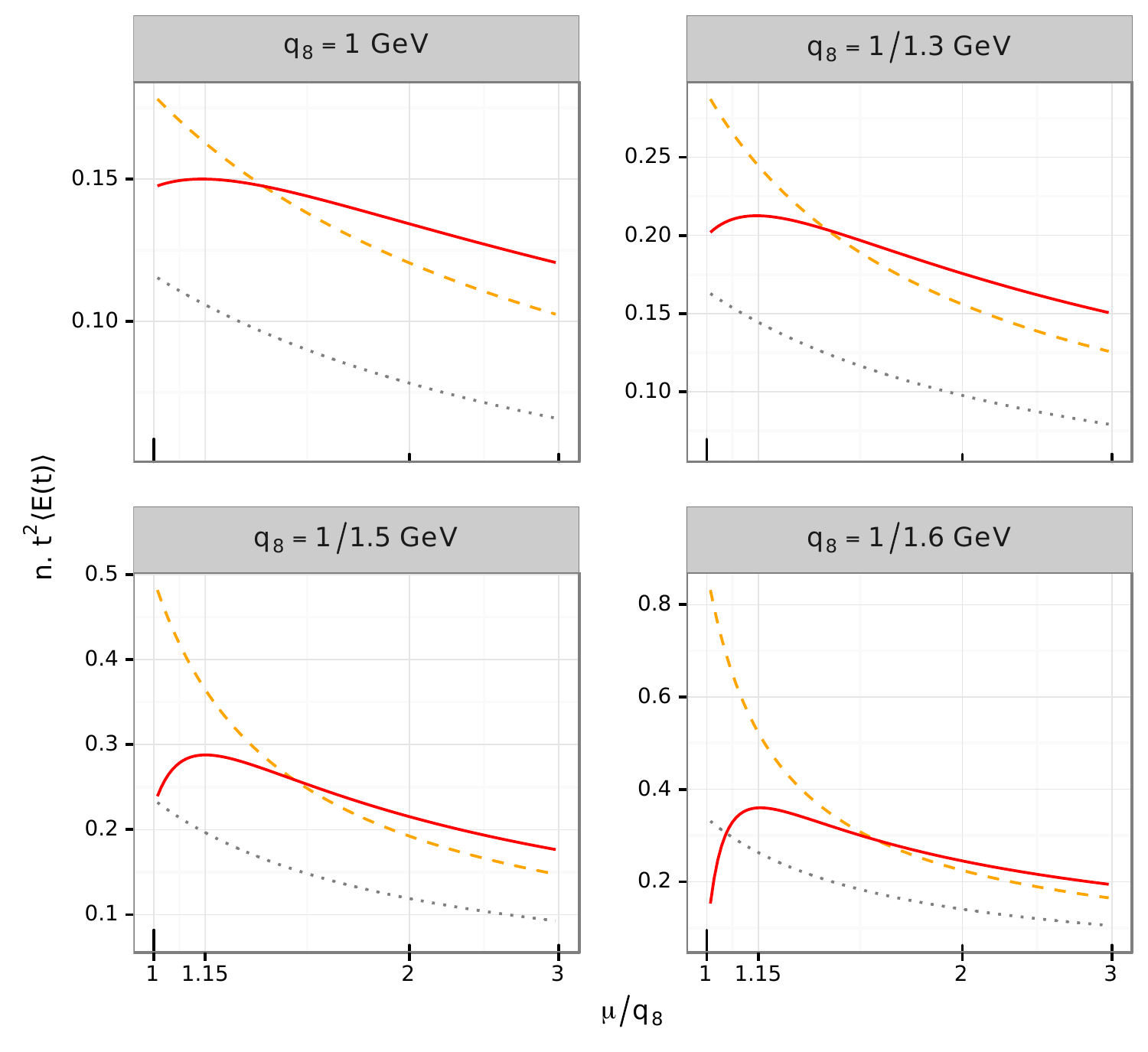}
  \begin{center}
    \parbox{.9\textwidth}{
      \caption[]{\label{fig:mudep}\sloppy Same as \fig{fig:mudep2}, but
        for lower values of $q_8$, and restricted to the interval
        $\mu\in[q_8,3q_8]$. }}
  \end{center}
\end{figure}

\fig{fig:mudep2} shows the dependence of $\tet$ as a function of
$\mu/q_8$ for $q_8=100$\,GeV and $q_8=2$\,GeV.  In both cases, one
observes a sound perturbative behavior in the interval
$\mu\in[q_8,3q_8]$. In addition, the $\mu$-dependence decreases
significantly with increasing loop order. These features quickly fade
away when going to lower values of $\mu$. Our conclusion is that the
best prediction for $\tet$ is obtained within the $\mu$-interval
$[q_8,3q_8]$; its variation within this interval will be used as an
estimate of the theoretical uncertainty. Values of $\mu$ outside this
interval will be disregarded in what follows.

\fig{fig:mudep} shows $\tet$ within this interval for a few values of
$q_8\leq 1$\,GeV.  It is interesting to note that for $q_8=1/1.5$\,GeV,
corresponding to $\sqrt{t}\approx 0.1$\,fm, we may still make
quantitative predictions when focussing on the $\mu$-interval identified
above. For lower energies, the uncertainty at \nnlo{} becomes of the
order of 100\%, and the \nlo{} and \nnlo{} correction are of the same
order of magnitude.

A common feature of all the plots in \figs{fig:mudep2} and
\ref{fig:mudep} (except the one at $q_8=1/1.6$\,GeV, a value which we
will not consider any further in this paper) is that, within
$\mu\in[q_8,3 q_8]$, the maximum is quite precisely at $\mu=1.15\,q_8$,
while the minimum is at $\mu=3q_8$.  Therefore, the error interval of
$\tet$ as defined above is given to a very good approximation by its
values at $\mu=\mu_-\equiv 3q_8$ and $\mu=\mu_+\equiv 1.15q_8$.

\begin{figure}
  \begin{center}
    \includegraphics[width=.8\textwidth]{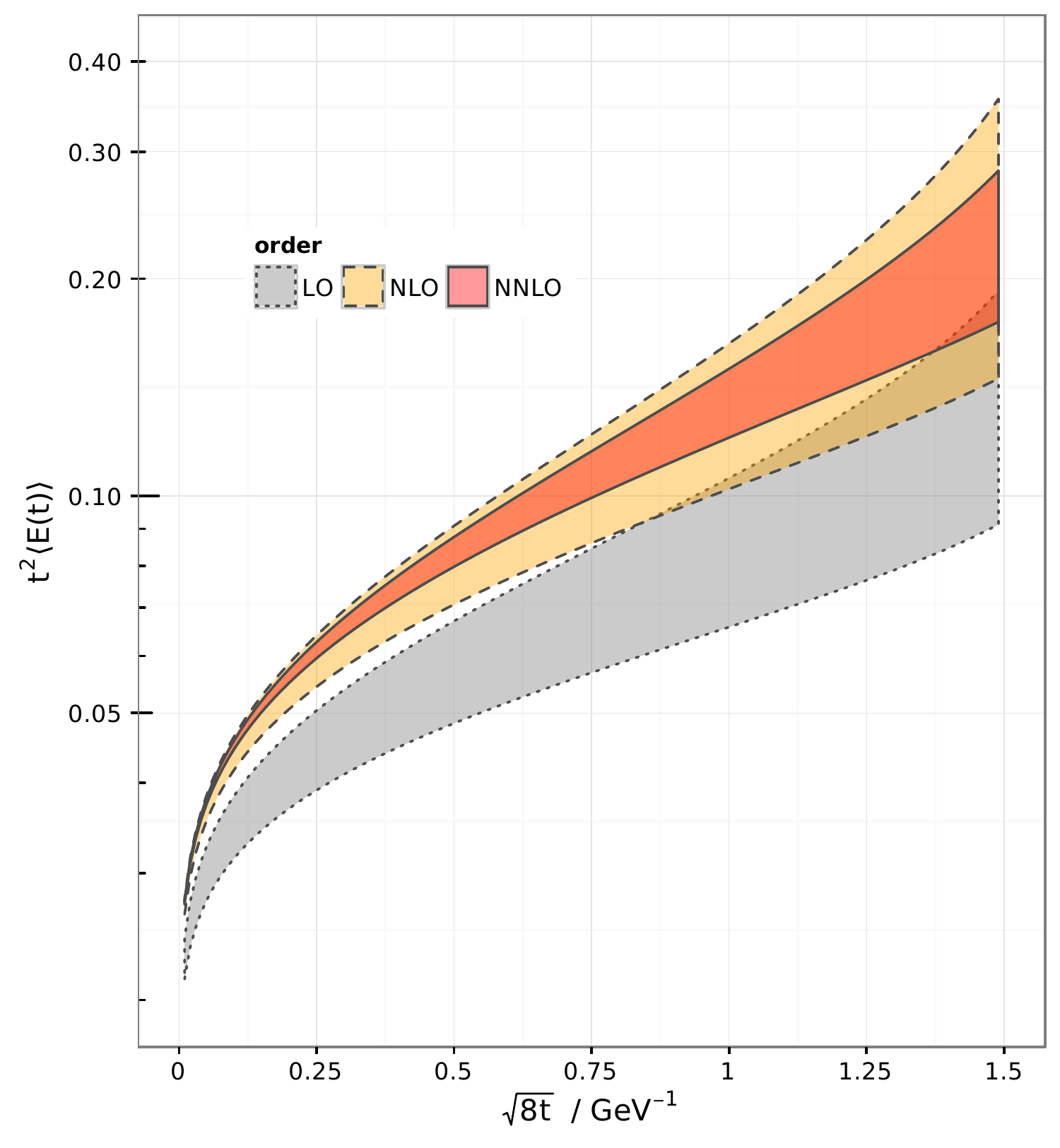}
    \parbox{.9\textwidth}{
      \caption[]{\label{fig:et}\sloppy $\tet$ for
        $n_f=3$ as a function of $\sqrt{8t}$ (in GeV$^{-1}$) for
        $\mu=3/\sqrt{8t}$ (lower) and $\mu=1.15/\sqrt{8t}$ (upper) at
        \lo{} (gray), \nlo{} (orange), and \nnlo{} (red).}}
  \end{center}
\end{figure}

\fig{fig:et} shows the dependence of $\tet$ on $\sqrt{8t}=1/q_8$ for
$n_f=3$ active flavors at \lo{}, \nlo{}, and \nnlo{}, with error bands
evaluated as indicated above. For each value of $q_8$, the strong
coupling $\as$ is evolved at four-loop level from $\alpha_s^{(5)}(\mz)$
to $\alpha_s^{(3)}(q_8)$ (including three-loop matching at the quark
thresholds), and subsequently at the pertinent order from
$\as^{(3)}(q_8)$ to $\as^{(3)}(\mu)$, with $\mu=1.15 q_8$ and $\mu=3q_8$
for the upper and lower edge of the uncertainty band, respectively. One
observes that the resulting \nlo{} and the \nnlo{} bands nicely overlap,
which gives confidence in using these bands as measures of the
theoretical uncertainty. There is hardly any overlap of these curves
with the \lo{} band though.

\subsection{Extracting $\as(\mz)$}\label{sec:alphas}

One of the most interesting applications of our results would be the
derivation of a numerical value of $\as(\mz)\equiv \alpha_s^{(5)}(\mz)$
using lattice data as input. This will be most promising, of course, if
the lattice calculation for $\tet$ could be extended to the perturbative
regime, which seems to have become a realistic
perspective\,\cite{Luscher:2014kea}.

Assume that a lattice value $e(t)$ for $\tet$ is known, evaluated at
$t=1/(8q_8^2)$ and for $n_f$ active quark flavors. Using the
perturbative result of \eqns{eq:nnlotet} and \noeqn{eq:nnloke} through
order $l$ (including its $\mu$-dependence), one can derive an $l$-loop
value for $\as^{(n_f)}(\mu)$, which can then be converted into a value
for $\as^{(5)}(\mz)$ through four-loop {\abbrev RG} evolution and
three-loop matching to the $n_f=5$ theory.  Table\,\ref{tab:etfromas}
shows this relation at \nnlo{} (i.e.\ for $l=3$) for a number of values
of $q_8$ and $n_f$. The values of $\tet$ given in that table correspond
to the center of the error band, i.e., they are the arithmetic means of
$\tet$ evaluated at $\mu=1.15\,q_8$ and $\mu=3\,q_8$.  These numbers take
into account the \nlo{} quark effects given in \eqn{eq:omega1q},
whereupon the lightest three quark flavors are taken massless, while
$m_c=1.67$\,GeV and $m_b=4.78$\,GeV. The mass effects therefore only
affect the columns with $n_f\geq 4$. At $q_8=2$\,GeV, their effect on
$\tet$ is about 0.8\%, at $q_8=10$\,GeV it is less than 0.3\% both for
$n_f=4$ and $n_f=5$, while at $q_8=\mz$, they have no effect on the
digits given in the table.

\begin{table}
\begin{center}
\begin{tabular}{|l||c|c|c|c|c|c|c|c|}
\hline
&\multicolumn{8}{|c|}{$t^2\langle E(t)\rangle\cdot 10^4$}\\
\hline
$q_8$&\multicolumn{2}{c|}{2\,GeV} &\multicolumn{3}{c|}{10\,GeV} &\multicolumn{3}{c|}{$m_Z$}\\
\hline
$\alpha_s(m_Z)$  & $n_f=3$  & $n_f=4$  & $n_f=3$  & $n_f=4$  & $n_f=5$  & $n_f=3$  & $n_f=4$  & $n_f=5$ \\
\hline
0.113 & 744 & 755 & 424 & 446 & 456 & 267 & 285 & 299 \\ 
0.1135 & 753 & 764 & 426 & 449 & 459 & 268 & 286 & 301 \\ 
0.114 & 762 & 773 & 429 & 452 & 462 & 269 & 287 & 302 \\ 
0.1145 & 771 & 782 & 432 & 455 & 466 & 270 & 289 & 303 \\ 
0.115 & 780 & 792 & 435 & 458 & 469 & 272 & 290 & 305 \\ 
0.1155 & 789 & 802 & 438 & 461 & 472 & 273 & 291 & 306 \\ 
0.116 & 798 & 811 & 440 & 465 & 476 & 274 & 292 & 308 \\ 
0.1165 & 808 & 821 & 443 & 468 & 479 & 275 & 294 & 309 \\ 
0.117 & 818 & 832 & 446 & 471 & 483 & 276 & 295 & 311 \\ 
0.1175 & 827 & 842 & 449 & 474 & 486 & 277 & 296 & 312 \\ 
0.118 & 837 & 852 & 452 & 478 & 490 & 278 & 298 & 314 \\ 
0.1185 & 847 & 863 & 455 & 481 & 493 & 279 & 299 & 315 \\ 
0.119 & 858 & 874 & 457 & 484 & 497 & 280 & 300 & 316 \\ 
0.1195 & 868 & 885 & 460 & 488 & 500 & 281 & 301 & 318 \\ 
0.12 & 879 & 896 & 463 & 491 & 504 & 282 & 303 & 319 \\ 
\hline
\end{tabular}

\caption[]{\label{tab:etfromas}Numerical values for $10^4\cdot\tet$
  corresponding to various $\as(\mz)\equiv \as^{(5)}(\mz)$. Given a
  numerical result for $\tet$ (e.g., from a lattice calculation), this
  table lets one deduce the corresponding value of $\as(\mz)$. The
  associated perturbative uncertainty for $n_f=5$ and $n_f=3$ can be
  read off from \fig{fig:plot_as}.}
\end{center}
\end{table}

\begin{figure}
  \begin{center}
    \includegraphics[width=.8\textwidth]{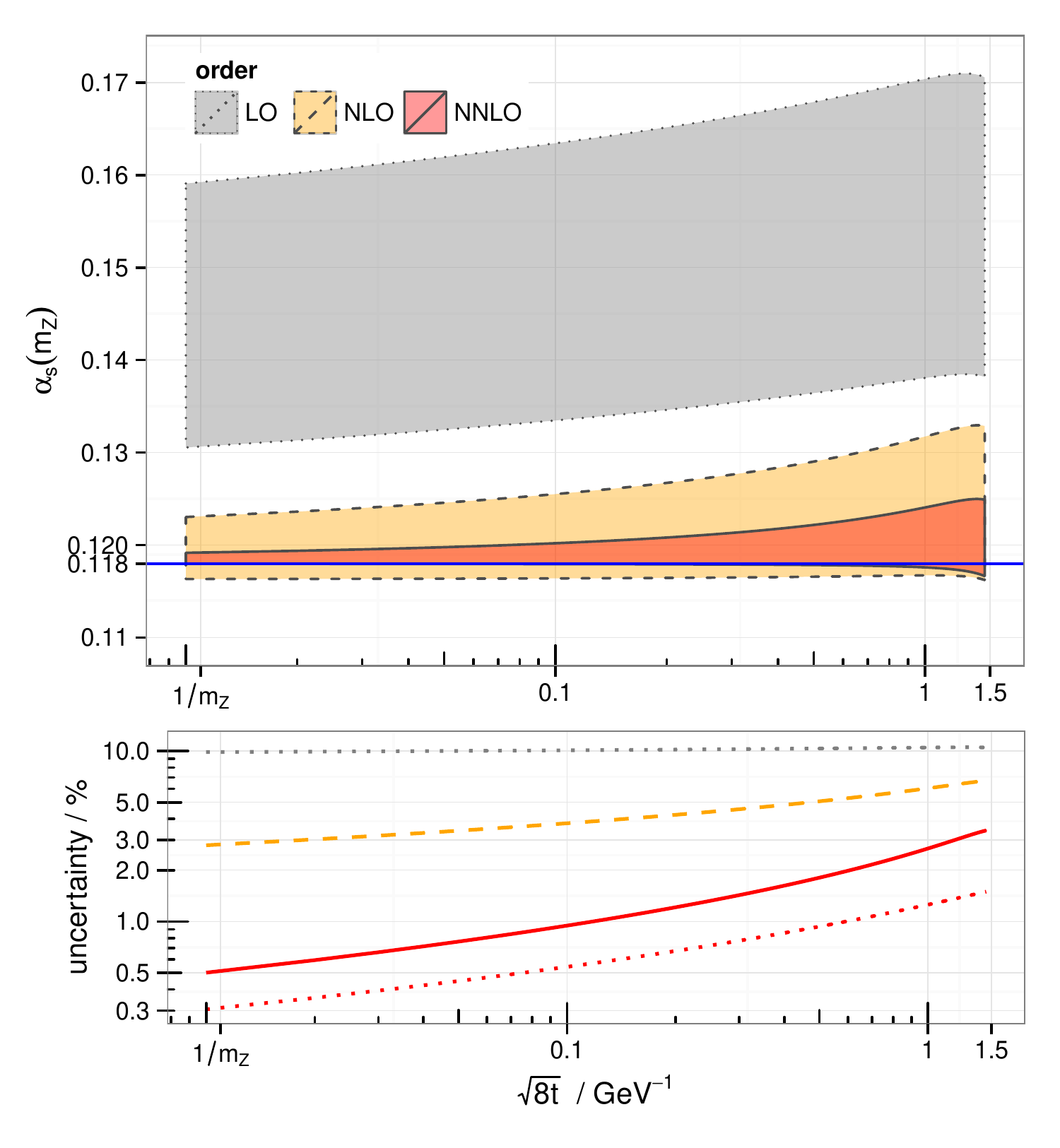}
    \parbox{.9\textwidth}{
      \caption[]{\label{fig:plot_as} Upper plot: numerical value for
        $\as^{(5)}(\mz)$ derived at \lo{} (gray), \nlo{} (orange), and
        \nnlo{} (red) from a hypothetical exact value of $\tet|_{n_f=3}$
        (see main text for details). Lower plot: corresponding
        theoretical uncertainty (see \eqn{eq:uncertainty}). The red
        dotted line in the lower plot shows the uncertainty when the
        analysis is based on $\tet|_{n_f=5}$.}}
  \end{center}
\end{figure}

In accordance with our previous considerations, we estimate the
theoretical accuracy of this extraction by considering $\tet$ at
$\mu=1.15\, q_8$ and $3\,q_8$ when deriving $\as^{(n_f)}(\mu)$ from
$e(t)$.  The result for $n_f=3$ is shown in \fig{fig:plot_as}. In lack
of a precise value of $e(t)$ at sufficiently large $q_8$, we substitute
it by the perturbative \nnlo{} expression for $\tet$ at $\mu=q_8$, where
the numerical value for $\as^{(3)}(q_8)$ is derived by three-loop
running and two-loop matching ($\mu_b=m_b$ and $\mu_c=2m_c$) from the
input value $\alpha_s(\mz)=0.118$. Therefore, the \nnlo{} band for
$\as(\mz)$ in the upper part of \fig{fig:plot_as} always includes the
value 0.118 by construction. Similar to \fig{fig:et}, the width of the
bands decreases remarkably towards higher orders of perturbation theory.
The \nnlo{} band lies completely within the \nlo{} band, while \lo{} has
no overlap with \nlo{}.

The lower part of the figure shows the theoretical accuracy that could
be achieved by such an analysis, derived by taking the relative width of
the bands of the upper part of the plot, 
\begin{equation}
\begin{split}
\frac{\Delta\as}{\as} = \frac{\as^\text{max}(\mz) -
  \as^\text{min}(\mz)}{\as^\text{max}(\mz) +
  \as^\text{min}(\mz)}\,.
\label{eq:uncertainty}
\end{split}
\end{equation}
For example, if $e(t)$ is given only at $t=1/(8 \text{GeV}^2)$, the
\nnlo{} uncertainty on $\as(\mz)$ would be around $2.5\%$. On the other
hand, knowning $e(t)$ at $t=1/(8\mz^2)$ would allow one to derive
$\as(\mz)$ to $0.5\%$ accuracy which is at the same level as the
current world average on this quantity\,\cite{Agashe:2014kda}. Also
shown in the lower plot is the uncertainty which results from knowing
$e(t)$ for $n_f=5$ active flavors (lower dotted red line). In this case,
the numbers above decrease to $\sim 1.1\%$ and $\sim 0.3\%$,
respectively, because of the lower value of the \qcd{} $\beta$ function.

\subsection{Derivative of the action density}\label{sec:deriv}

In \citere{Borsanyi:2012zs} it was argued that the quantity
\begin{equation}
\begin{split}
W(t)\equiv t\frac{\dd}{\dd t}\tet
\end{split}
\end{equation}
is more suitable for scale setting on the lattice.  Neglecting again
quark-mass effects, $t$ and $\mu$ are the only dimensional scales of the
dimensionless quantity $\tet$, so that the dependence on them can only
be in terms of $l_{t\mu} \equiv \ln t\mu^2$. Using
\eqn{eq:rginv}, we can thus write
\begin{equation}
\begin{split}
W(t) = \frac{\partial}{\partial l_{t\mu}}\tet = -
\as\beta(\as)\frac{\partial}{\partial\as}\tet\,,
\end{split}
\end{equation}
with the $\beta$ function defined in \eqn{eq:rgeq}. The result is
therefore
\begin{equation}
\begin{split}
W(t) &= \frac{3}{4}\left(\api\right)^2\beta_0\left[
1 + \as\left(b_1 + 2k_1\right) + \as^2\left(b_2+2\,b_1
k_1 + 3 k_2\right)\right]\,,
\end{split}
\end{equation}
with $k_1$, $k_2$ given in \eqn{eq:nnloke}, $b_n\equiv
\beta_n/(\pi^n\beta_0)$, where $\beta_0$ and $\beta_1$ have been given
in \eqn{eq:beta}, and\footnote{Again, we give only the \qcd{} expression
  here. For the coefficient in a general Lie group, see
  \citere{vanRitbergen:1997va,Czakon:2004bu}.}
\begin{equation}
\begin{split}
\beta_2 = \frac{2857}{128} - \frac{5033}{1152}\,n_f + \frac{325}{3456}\,n_f^2\,.
\end{split}
\end{equation}
Numerically, this gives, for \qcd{} and setting $\mu=1/\sqrt{8t}$,
\begin{equation}
\begin{split}
W(t) &= \as^2\,\left(0.208975 - 0.0126651\,n_f\right) 
\\&+ \as^3\left(0.613022 - 0.0437989\,n_f - 
   0.000191375\,n_f^2\right) 
\\&+
 \as^4\,\left(-0.10538(3) - 0.0798618(4)\,n_f + 
   0.00426484(9)\,n_f^2 - 0.0000711364\,n_f^3\right)\,.
\end{split}
\end{equation}
Again, the $\mu$-dependent terms can be easily reconstructed using
renormalization group invariance. 

Performing a similar analysis for $W(t)$ as done in the preceeding
sections for $\tet$, we see no improvement concerning the precision for
the extraction of $\as$ relative to the one based on $\tet$.

\section{Conclusions and Outlook}\label{sec:conclusions}

The action density for \qcd{} gradient flow fields has been evaluated at
three-loop level. The perturbative expansion has been derived by
standard Wick contractions, and the resulting integrals have been solved
by sector decomposition supplied by a suitable numerical integration
algorithm. A number of strong checks on the result has been
performed. In addition, quark-mass effect have been included at \nlo{}.

Our \nnlo{} coefficient indicates a very well-behaved perturbative
series for the action density down to energy scales of about $q_8 \sim
0.65$\,GeV, corresponding to $\sqrt{t} \sim 0.11$\,fm.  This seems well
within reach of a direct comparison to a lattice evaluation of $\tet$.
Given that $\tet$ can be evaluated independently (e.g.\ by a lattice
calculation) at sufficiently large values of the flow time with high
precision, one may derive a numerical value for $\as(\mz)$ by
comparison to the perturbative result. We provide an estimate of the
resulting uncertainty and find that it could be competitive with the
current world average.

On the perturbative side, further steps could be the development of more
efficient tools for the evaluation of the integrals, the consideration
of other observables, or the application of the flow-field formalism to
quark fields as introduced in \citere{Luscher:2013cpa}.

Finally, it should be noted that there is no conceptual limitation of
the calculational method described in this paper which would restrict it
to the three-loop level. In the current implementation, however, an
extension to four loops would require a significant increase in the
computing resources.

\paragraph{Acknowledgments.}

We are particularly obliged to Zoltan Fodor for initiating and
motivating this work, to Martin L\"uscher for providing us with private
notes on \citere{Luscher:2010iy}, to Szabolcs Bors\'anyi, Christian
H\"olbling, and Rainer Sommer for helpful communication, and to Marisa
Sandhoff and Torsten Harenberg for administration of the {\abbrev DFG
  FUGG} cluster at Bergische Universit\"at Wuppertal, where most of the
calculations for this paper were performed.

\def\app#1#2#3{{\it Act.~Phys.~Pol.~}\jref{\bf B #1}{#2}{#3}}
\def\apa#1#2#3{{\it Act.~Phys.~Austr.~}\jref{\bf#1}{#2}{#3}}
\def\annphys#1#2#3{{\it Ann.~Phys.~}\jref{\bf #1}{#2}{#3}}
\def\cmp#1#2#3{{\it Comm.~Math.~Phys.~}\jref{\bf #1}{#2}{#3}}
\def\cpc#1#2#3{{\it Comp.~Phys.~Commun.~}\jref{\bf #1}{#2}{#3}}
\def\epjc#1#2#3{{\it Eur.\ Phys.\ J.\ }\jref{\bf C #1}{#2}{#3}}
\def\fortp#1#2#3{{\it Fortschr.~Phys.~}\jref{\bf#1}{#2}{#3}}
\def\ijmpc#1#2#3{{\it Int.~J.~Mod.~Phys.~}\jref{\bf C #1}{#2}{#3}}
\def\ijmpa#1#2#3{{\it Int.~J.~Mod.~Phys.~}\jref{\bf A #1}{#2}{#3}}
\def\jcp#1#2#3{{\it J.~Comp.~Phys.~}\jref{\bf #1}{#2}{#3}}
\def\jetp#1#2#3{{\it JETP~Lett.~}\jref{\bf #1}{#2}{#3}}
\def\jphysg#1#2#3{{\small\it J.~Phys.~G~}\jref{\bf #1}{#2}{#3}}
\def\jhep#1#2#3{{\small\it JHEP~}\jref{\bf #1}{#2}{#3}}
\def\mpl#1#2#3{{\it Mod.~Phys.~Lett.~}\jref{\bf A #1}{#2}{#3}}
\def\nima#1#2#3{{\it Nucl.~Inst.~Meth.~}\jref{\bf A #1}{#2}{#3}}
\def\npb#1#2#3{{\it Nucl.~Phys.~}\jref{\bf B #1}{#2}{#3}}
\def\nca#1#2#3{{\it Nuovo~Cim.~}\jref{\bf #1A}{#2}{#3}}
\def\plb#1#2#3{{\it Phys.~Lett.~}\jref{\bf B #1}{#2}{#3}}
\def\prc#1#2#3{{\it Phys.~Reports }\jref{\bf #1}{#2}{#3}}
\def\prd#1#2#3{{\it Phys.~Rev.~}\jref{\bf D #1}{#2}{#3}}
\def\pR#1#2#3{{\it Phys.~Rev.~}\jref{\bf #1}{#2}{#3}}
\def\prl#1#2#3{{\it Phys.~Rev.~Lett.~}\jref{\bf #1}{#2}{#3}}
\def\pr#1#2#3{{\it Phys.~Reports }\jref{\bf #1}{#2}{#3}}
\def\ptp#1#2#3{{\it Prog.~Theor.~Phys.~}\jref{\bf #1}{#2}{#3}}
\def\ppnp#1#2#3{{\it Prog.~Part.~Nucl.~Phys.~}\jref{\bf #1}{#2}{#3}}
\def\rmp#1#2#3{{\it Rev.~Mod.~Phys.~}\jref{\bf #1}{#2}{#3}}
\def\sovnp#1#2#3{{\it Sov.~J.~Nucl.~Phys.~}\jref{\bf #1}{#2}{#3}}
\def\sovus#1#2#3{{\it Sov.~Phys.~Usp.~}\jref{\bf #1}{#2}{#3}}
\def\tmf#1#2#3{{\it Teor.~Mat.~Fiz.~}\jref{\bf #1}{#2}{#3}}
\def\tmp#1#2#3{{\it Theor.~Math.~Phys.~}\jref{\bf #1}{#2}{#3}}
\def\yadfiz#1#2#3{{\it Yad.~Fiz.~}\jref{\bf #1}{#2}{#3}}
\def\zpc#1#2#3{{\it Z.~Phys.~}\jref{\bf C #1}{#2}{#3}}
\def\ibid#1#2#3{{ibid.~}\jref{\bf #1}{#2}{#3}}
\def\otherjournal#1#2#3#4{{\it #1}\jref{\bf #2}{#3}{#4}}
\newcommand{\jref}[3]{{\bf #1} (#2) #3}
\newcommand{\hepph}[1]{\href{http://arXiv.org/abs/hep-ph/#1}{{\tt hep-ph/#1}}}
\newcommand{\hepth}[1]{\href{http://arXiv.org/abs/hep-th/#1}{{\tt hep-th/#1}}}
\newcommand{\heplat}[1]{\href{http://arXiv.org/abs/hep-lat/#1}{{\tt hep-lat/#1}}}
\newcommand{\mathph}[1]{\href{http://arXiv.org/abs/math-ph/#1}{{\tt
      math-ph/#1}}}
\newcommand{\arxiv}[2]{\href{http://arXiv.org/abs/#1}{{\tt arXiv:#1}}}
\newcommand{\bibentry}[4]{#1, {\it #2}, #3\ifthenelse{\equal{#4}{}}{}{, }#4.}

\end{document}